\documentclass[aps,pre,reprint,groupedaddress]{revtex4-1}

\usepackage{graphicx}
\usepackage{amssymb}
\usepackage{amsmath}
\usepackage{bm}
\usepackage{hyperref}
\usepackage{lipsum}
\usepackage{bbold}
\usepackage{comment}
\usepackage{color}

\usepackage{babel}

\newcommand\cmoire{Moire~}
\newcommand\lmoire{moire~}

\begin{document}

\title{Stacking and gate tunable topological flat bands, gaps and anisotropic strip patterns in twisted trilayer graphene}

\author{Jiseon \surname{Shin}} 
\affiliation{Department of Physics, University of Seoul, Seoul 02504, Korea}
\author{Bheema Lingam \surname{Chittari}}
\affiliation{Department of Physical Sciences, Indian Institute of Science Education and Research Kolkata, Mohanpur 741246, West Bengal, India}
\author{Jeil \surname{Jung}}
\email[jeiljung@uos.ac.kr]{}
\affiliation{Department of Physics, University of Seoul, Seoul 02504, Korea}
\affiliation{Department of Smart Cities, University of Seoul, Seoul 02504, Korea}

\date[]{}

\begin{abstract}
Trilayer graphene with a twisted middle layer has recently emerged as a new platform exhibiting 
correlated phases and superconductivity near its magic angle. 
A detailed characterization of its electronic structure in the parameter space of twist angle $\theta$, 
interlayer potential difference $\Delta$, and top-bottom layer stacking $\bm{\tau}$ reveals that
flat bands with large Coulomb energy vs bandwidth $U/W > 1$ 
are expected within a range of $\pm 0.2^{\circ}$ near $\theta \simeq1.5^{\circ}$ 
and $\theta \simeq1.2^{\circ}$ for $\bm{\tau}_{\rm AA}$ top-bottom layer stacking,
between a wider $1^{\circ} \sim 1.7^{\circ}$ range for $\bm{\tau}_{\rm AB}$ stacking, 
whose bands often have finite valley Chern numbers thanks to the opening of primary and secondary band gaps in the presence of a finite $\Delta$,
and below $\theta \lesssim 0.6^{\circ}$ for all $\bm{\tau}$ considered. 
The largest $U/W$ ratios are expected at the magic angle $\sim 1.5^{\circ}$ when $|\Delta| \sim 0$~meV for AA, 
and slightly below near $\sim 1.4^{\circ}$ for finite $|\Delta| \sim 25$~meV for AB stackings, 
and near $\theta \sim 0.4^{\circ}$ for both stackings.
%
When $\bm{\tau}$ is the saddle point stacking vector between AB and BA we observe 
pronounced anisotropic local density of states (LDOS) strip patterns with broken triangular rotational symmetry.
We present optical conductivity calculations that reflect the changes in the electronic structure 
introduced by the stacking and gate tunable system parameters. 
\end{abstract}

\maketitle

\section{Introduction}
Research on the electronic structure of nearly flat bands in moire materials has seen a recent surge of interest following experimental 
observation of strongly correlated and localized Mott-like phases 
and superconductivity in magic angle twisted bilayer graphene (tBG)~\cite{Kim2017, Cao2018a, Cao2018b, Yankowitz2018a, Cao2019a}
discussed by electronic structure studies~\cite{SuarezMorell2010, Bistritzer2010a,Jung2014}.
Research interests in vertical van der Waals (vdW) heterojunctions~\cite{Koma1992, Koma1999, Geim2013a} 
in search of strongly correlated flat bands have expanded beyond the twisted bilayer graphene~\cite{Bistritzer2010a, SuarezMorell2010, Kim2017, Cao2018a, Cao2018b, Yankowitz2018a, Koshino2018, Cao2019a, Leconte2019, Koshino2019a}, to include systems like twisted double bilayer graphene (tDBG)~\cite{Chebrolu2019, Koshino2019, Burg2019, Lee2019a, Choi2019a}, and various forms of 
twisted trilayer graphene~\cite{SuarezMorell2013, Zuo2018, Ma2019,Mora2019, Tsai2019, Khalaf2019,Li2019,Carr2019a,szendr2020ultraflat,
tmbg1,tmbg2,tmbg3,Park2020, lei2020mirror, wu2020lattice, Dumitru2021} including twisted monolayer-bilayer graphene (tMBG)~\cite{SuarezMorell2013,Li2019,szendr2020ultraflat,Carr2019a,Ma2019,
tmbg1,tmbg2,tmbg3,Park2020, lei2020mirror, wu2020lattice}.
Unlike the systems with a single moire twist interface like tBG, tDBG, or tMBG,
in twisted trilayer graphene with finite successive interlayer twist angles
we have two interfaces giving rise to double moire patterns. 
When these moire patterns are mutually incommensurate they give rise to supermoire patterns~\cite{supermoire1,supermoire2,supermoire3},
also called moire of moire patterns~\cite{Kerelsky2019, Zhu2019, Tsai2019} that can multiply the features in the electronic structure, while strongest double moire interference happen for commensurate patterns,
exemplified by the large secondary band gaps in graphene encapsulated by 
hexagonal boron nitride~\cite{supermoire3}.

Commensurate double moire trilayer graphene with a middle layer twist~\cite{Ma2019,Mora2019,Khalaf2019, Carr2019a, Li2019, tb_tTG,lei2020mirror,park2020tunable, hao2020electric, Dumitru2021,dumitru2021tstg},
called here simply twisted trilayer graphene (tTG), has emerged as a system of renewed interest thanks to the observation of 
moire flat band superconductivity with a critical temperature higher than tBG.~\cite{park2020tunable, hao2020electric, cao2021large}
Earlier studies reported that the first magic-angle of tTG 
is larger by a factor $\sqrt{2}$ than that of tBG~\cite{Khalaf2019,Li2019,Carr2019a} and it was shown 
tight-binding models that nearly flat bands accompany the linear dispersions at the $\widetilde{K}$ point in the moire Brillouin zone (mBZ) that
persists even when out-of-plane lattice relaxation and perpendicular electric fields are present~\cite{Carr2019a, tb_tTG}. 
Other earlier work based on continuum models have analyzed the properties of tTG from various perspectives, 
including the band topology~\cite{Ma2019}, predominant metallic character~\cite{Mora2019},
hierarchy of magic angles~\cite{Khalaf2019}, symmetry analysis~\cite{dumitru2021tstg}.
It was noted that the band structures vary considerably depending on the relative stacking
vector $\bm{\tau}$ between top and bottom layers~\cite{Li2019,lei2020mirror} 
that in the presence of out of plane relaxations and electric fields 
shows metallic bands for AA while a band gap opens for AB stackings~\cite{park2020tunable, hao2020electric}. 
Due to the large parameter space of twist angles, electric fields and stacking possibilities earlier work have reported
the electronic structure for select system parameters. 

In this work we present new phase diagrams of the bandwidths and valley Chern numbers
of the low energy nearly flat bands in the continuous parameter space of twist angles $\theta$ and the interlayer 
potential difference $\Delta$ for different $\bm{\tau}$ stacking vectors between top and bottom layers.
Our detailed calculations show that it is possible to achieve nearly flat bands 
prone to strong correlations in a relatively wide range $\pm 0.2^{\circ}$ 
of twist angles around the $\theta = 1.5^{\circ}$ magic angle and around 
$\theta \simeq1.2^{\circ}$ in the presence of appropriate 
interlayer potential difference $\Delta$ for AA top-bottom layer stacking,
and in an even broader $1^{\circ} \sim 1.7^{\circ}$ range if the top and bottom layers 
stacking is AB (or equivalently BA) where a finite $\Delta$ can isolate the bands 
by opening primary and secondary gaps over a wide range or parameters 
often leading to finite valley Chern numbers.
Additionally, we show the impact of stacking and electric fields in the local density of states (LDOS) maps that 
can be measured through scanning tunneling probes, 
and we present linear optical conductivity calculations for select stacking arrangements 
as a means to distinguish different electronic structures. 
Anisotropic moire patterns can be obtained for top-bottom layer sliding vectors 
that break the triangular rotational symmetry and the stripe patterns are 
maximized for the saddle point (SP) stacking vector suggesting that this type of stripe phases
could be favored when the system is subject to uniaxial strains or 
to boundary conditions that alter the stacking dependent energy landscape. 

Our manuscript is organized as follows. In Sec. \ref{Model} we introduce the model Hamiltonian,
Sec. \ref{R1} is devoted to the discussion of the electronic band structures 
for different interlayer potential difference and stacking configurations,
in Sec. \ref{R2} we discuss the numerical results of effective Coulomb interaction for the two different stackings and 
the valley Chern numbers, in Sec. \ref{R3} we discuss the anisotropy of the LDOS for non-symmetric stackings,
in Sec. \ref{R4} we report the numerical results on the longitudinal linear optical conductivity,
and in Sec \ref{Summary} we summarize our work.

\section{Model Hamiltonian}\label{Model}

The Hamiltonian of tTG with twisted middle layer 
can be captured by twisting the top-bottom and middle layers in opposite senses. 
The continuum model Hamiltonian for the $K$ valley is
\begin{equation}
H_\textrm{tTG} (\theta) = \left(\begin{array}{ccc}
h^-_b & T_1(\bm{r}) & 0  \\
T_1^{\dagger}(\bm{r}) & h^+_m& T_2(\bm{r}) \\
0 & T_2^{\dagger}(\bm{r}) & h^-_t\\
\end{array}\right) + V,
\label{eq1}
\end{equation}
where $V$ = diag($-\Delta$, $-\Delta$, 0, 0, $+\Delta$, $+\Delta$) is a $6\times6$ matrix 
that captures the interlayer potential difference due to an external electric field 
where we assume the interlayer potentials of $\pm \Delta$ at the top and bottom layers.
\begin{figure}
\begin{center}
\includegraphics[width=0.45\textwidth]{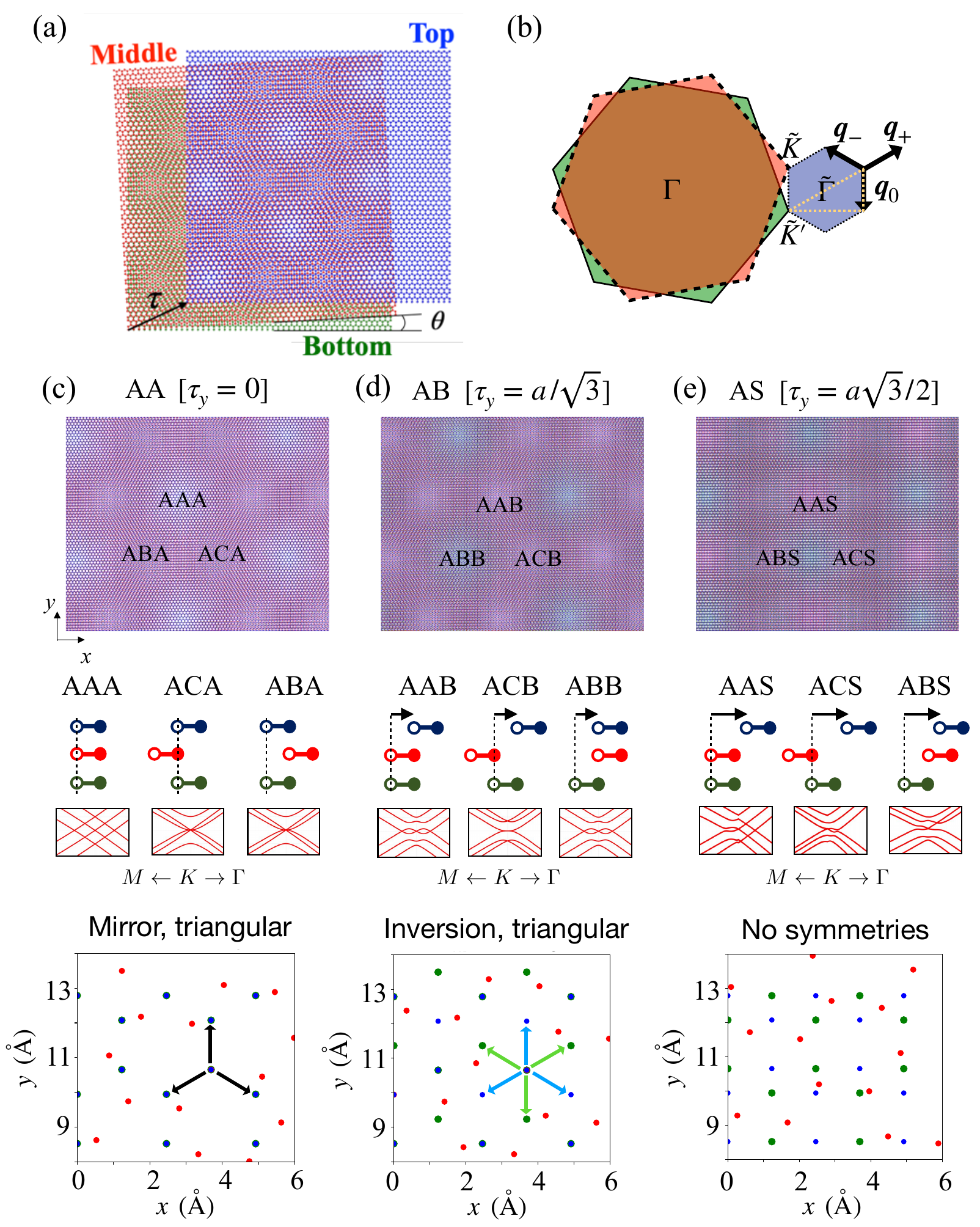}
\end{center}
\caption{
(Color online) (a) Schematic figure of the \lmoire patterns in tTG where the middle layer
is twisted by an angle $\theta$ and and the top layer slides by $\bm{\tau}$ with respect to the bottom layer. 
(b) Schematic figure for \lmoire Brillouin zone (mBZ) with its high-symmetry points. 
For $\theta = 1.5^{\circ}$ we have a \lmoire period of $\l_{\rm M} = 9.59$~nm. 
For (c) AA [$\tau_y = 0$], (d) AB [$\tau_y =a/\sqrt{3}$], 
and (e) AS-starting stacking [$\tau_y =a\sqrt{3}/2$], \cmoire patterns (upper row), 
corresponding commensurate stackings (middle row), and schematic diagram of atomic configurations 
at $\theta = 21.8^\circ$ are shown at the (bottom row). 
}\label{fig1}
\end{figure}

\begin{figure*}
\begin{center}
\includegraphics[width=1\textwidth]{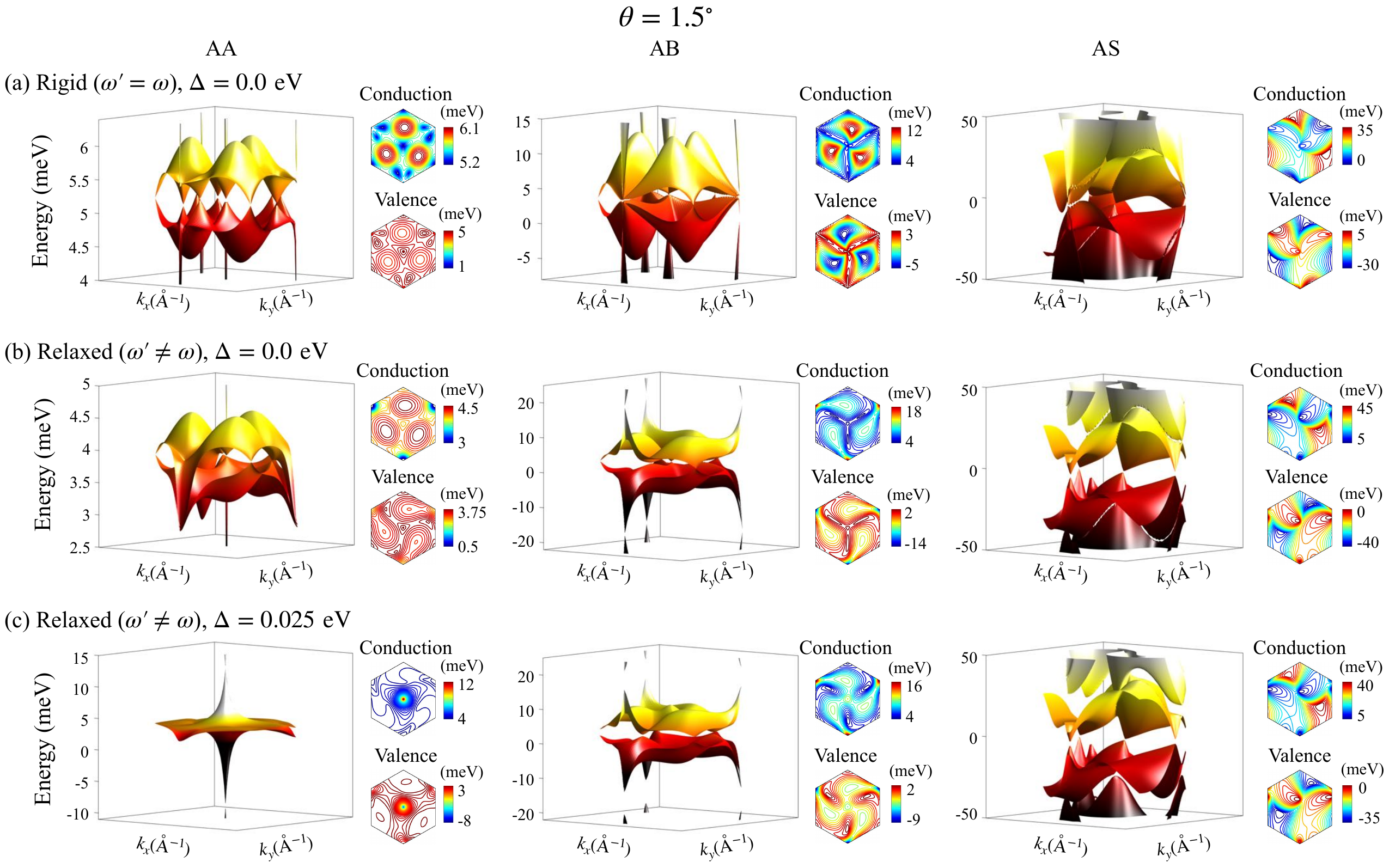}
\end{center}
\caption{
	Band structure of the low energy bands at the magic angle of $\theta =  1.5^{\circ}$ for three different AA, AB, AS stacking models that show a progressive bandwidth widening. 
	The interlayer tunneling are modeled for rigid $\omega' = \omega = 0.12$~eV and for out of plane relaxed by
	using unequal tunneling parameters $\omega' = 0.0939$~eV and $\omega = 0.12$~eV, which result in qualitative changes in the band structures and Fermi surface contours. 
	A finite interlayer potential difference $\Delta$ for AB stacked geometries widens the bandwidths and opens the band gaps 
	in the low energy bands. The anisotropic Fermi surface contours for the AS stacking reflects the broken
	triangular rotational symmetry noted also in the real space moire patterns. 
	}\label{fig2}
\end{figure*}

\begin{figure*}
\begin{center}
\includegraphics[width=1\textwidth]{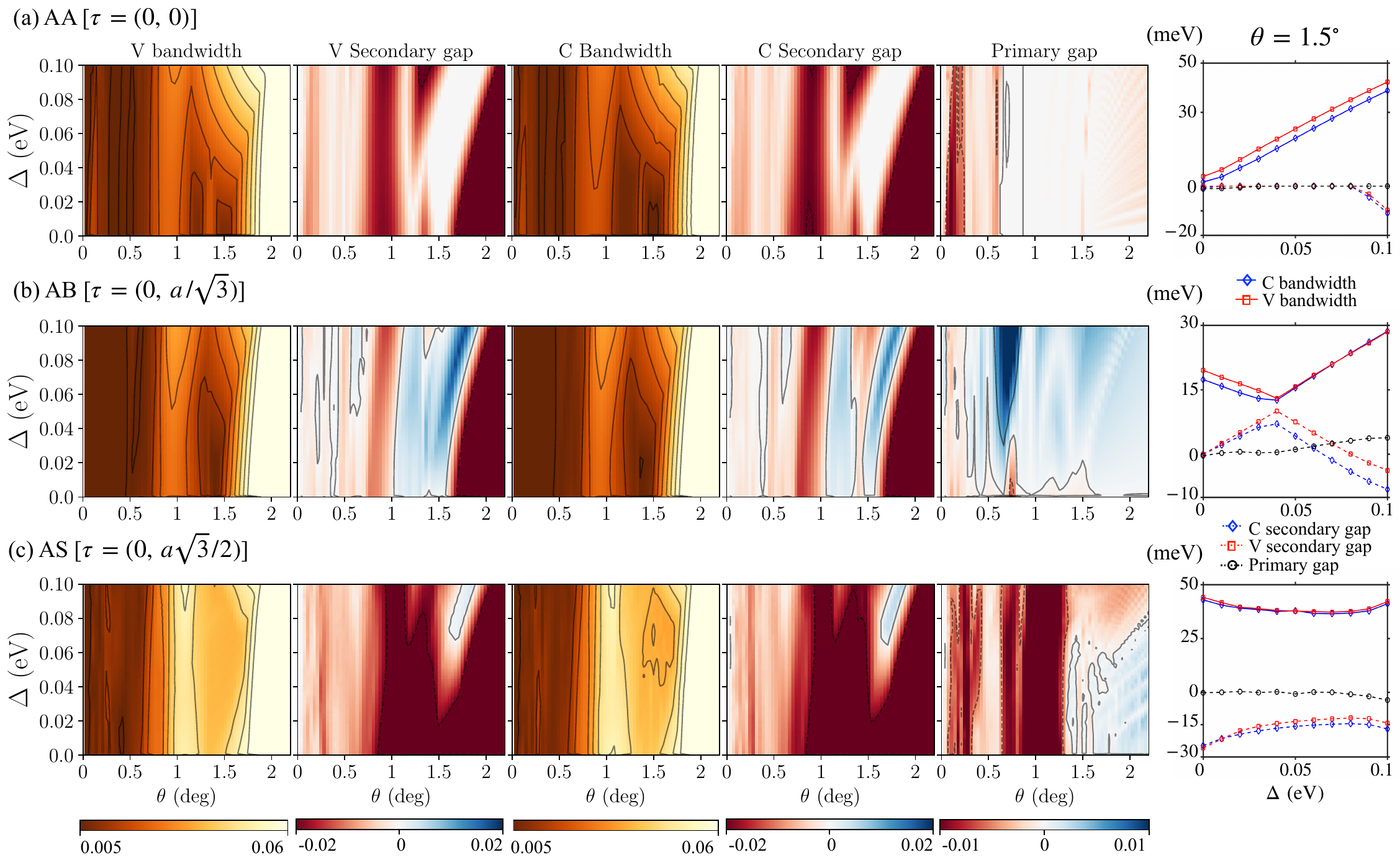}
\end{center}
\caption{
(Color online) The bandwidth,  secondary gap, primary gap of the conduction and valence bands as a 
function of the interlayer potential $\Delta$ and twist angle $\theta$ for (a) AA [$\tau_y = 0$], 
(b) AB [$\tau_y =a/\sqrt{3}$], and (c) SP stacking [$\tau_y =a\sqrt{3}/2$]. 
The one-dimensional cross sections of the bandwidths and bandgaps 
at $\theta = 1.5^\circ$ for the conduction and valence bands are presented at the rightmost column.
	}\label{fig3}
\end{figure*}
%
%

\begin{figure*}
\begin{center}
\includegraphics[width=0.9\textwidth]{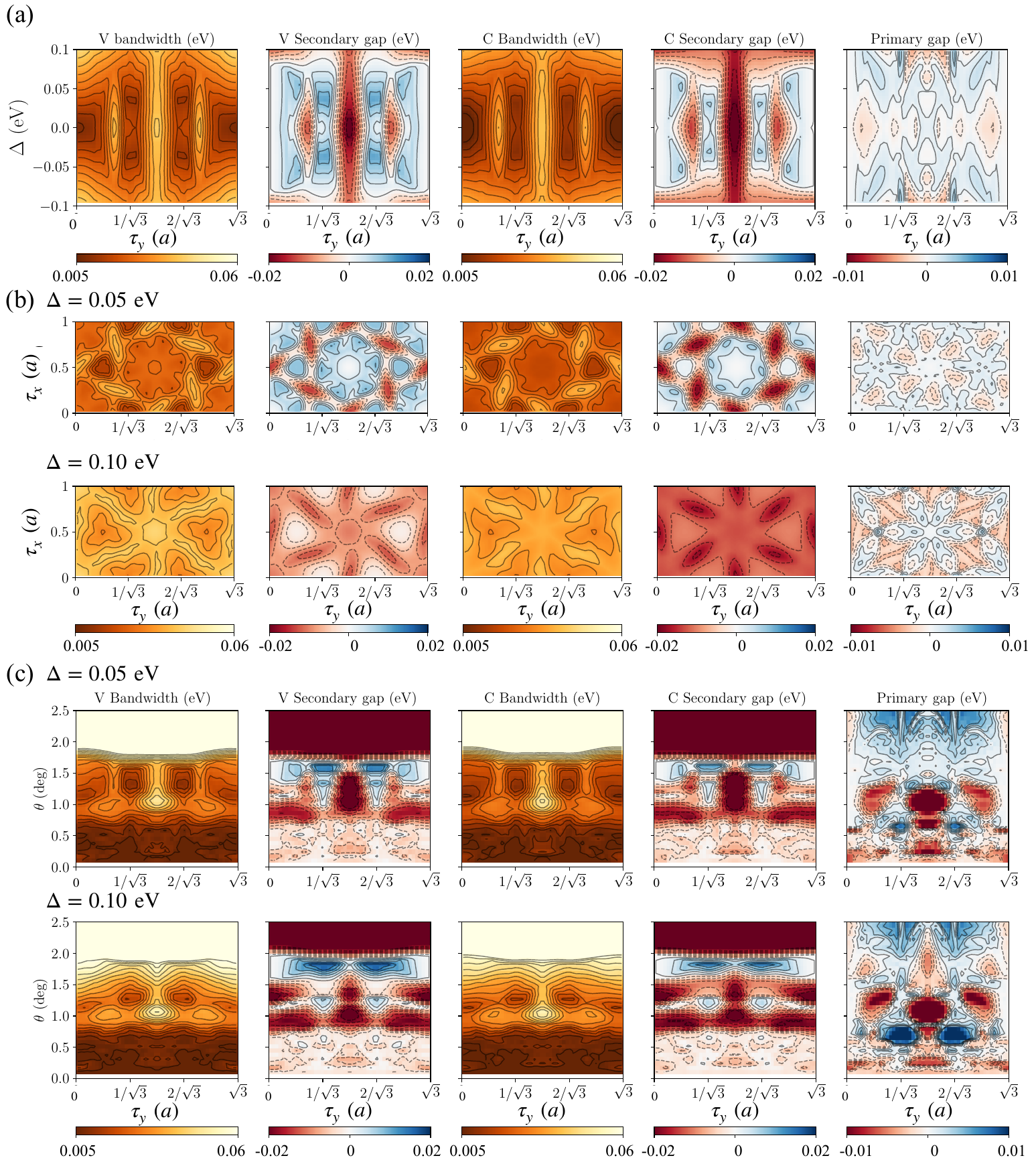}
\end{center}
\caption{ 
	(Color online) Bandwidths, secondary, and primary band gaps of the valence (V) and the conduction (C) bands at the magic angle $1.5^{\circ}$ (a) as a function of the interlayer potential difference ($\Delta$) and the sliding of the top G layer in $y$-direction $\tau_y$, (b) as a function of the sliding of the top G layer in $x$-, $y$-direction, $\tau_x$, $\tau_y$ for $\Delta = 0.05$~eV (upper row) and for $\Delta = 0.10$ eV (lower row), and (c) as a function of the twist angle $\theta$ and the sliding of the top G layer in the $y$-direction, 
$\tau_y$ for $\Delta = 0.05$ eV (upper row) and for $\Delta = 0.10$~eV (lower row).
	}\label{fig3b}
\end{figure*}

In our model, for the bottom (b), the middle (m), and the top (t) layers, $h^{\pm}_{b, m, t} = h(\pm \theta/2)$ represents a $2\times2$ matrix describing a Dirac Hamiltonian 
$h(\theta = 0) = v_F \bm{p} \cdot \bm{\sigma}$ rotated by $\pm \theta/2$ which is given as
\begin{equation}
h(\pm \theta/2)= D^\dagger (\pm \theta/2) ~ h(\theta = 0) ~ D (\pm \theta/2).
\end{equation}
The system conventions are similar to that in Ref.~\cite{Chebrolu2019} for tDBG and we 
use for the Fermi velocity $v_F = |t_0|\sqrt{3}a/2\hbar \simeq10^{6}$~m/s,
which corresponds to an effective nearest neighbor hopping term of $t_0 = -3.1$~eV. 
Here $D(\phi) = \exp(-i \phi \sigma_z /2)$ is in general a rotation operator for spin $S = 1/2$ and
$\sigma_z$ is the $z$-component of the Pauli matrices. 
The interlayer tunneling at the $i^{\rm th}$ interface is denoted as a $2\times2$ matrix $T_i(\bm{r})$ given by
\begin{equation}
T_k(\bm{r}) = \sum_{j=0, \pm} e^{i m_k \bm{q}_j \cdot \bm{r}} T^j_{s, s'}.
\end{equation}
where $m_k = (-1)^k$ and $\bm{q}_0$, $\bm{q}_\pm$ are given as $\bm{q}_0 = \theta k_D (0, -1)$, $ \bm{q}_\pm = \theta k_D (\pm \sqrt{3}/2, 1/2)$ when the twist angle $\theta$ is small enough. Here, $k_D = 4\pi/3 a$ is equal to the length of one side of the first Brillouin zone of single layer of graphene where $a = 2.46 \textrm{\AA}$. 
The interlayer tunning matrix $T^j_{s, s'}$ was first formulated for the local-AB-stacking in the twisted bilayer graphene 
system in Ref.~\cite{Bistritzer2010a} and was generalized for other initial stackings dictated by $\bm{\tau}$ in Ref.~\cite{Jung2014} as 
\begin{equation}
T^j =  e^{-i \bm{G}_j \cdot \bm{\tau}} \left(\begin{array}{cc}
\omega_{A,A'} & \omega_{A,B'} e^{-ij \varphi}  \\
\omega_{B,A'} e^{ij \varphi} & \omega_{B,B'} \\
\end{array}\right),
\label{eq_t}
\end{equation}
where $\bm{G}_0 = (0, 0)$ and $\bm{G}_\pm = k_D (-3/2, \pm \sqrt{3}/2)$. Here, $\bm{\tau} = (\tau_x,~\tau_y)$ is a relative sliding of the top layer with respect to the bottom layer, 
and we define $\omega' = \omega_{A,A'} = \omega_{B,B'}$, and $\omega = \omega_{A,B'} = \omega_{B,A'} $, resulting in 
\begin{equation}
T^0 = \left(\begin{array}{cc}
\omega' & \omega\\
\omega & \omega' \\
\end{array}\right), \hspace{0.3 cm} T^{\pm} = \left(\begin{array}{cc}
\omega' & \omega e^{\mp i\varphi}\\
\omega e^{\pm i\varphi} & \omega' \\
\end{array}\right),  \label{tunneling}
\end{equation}
when $\bm{\tau} = (\tau_x, \tau_y) = (0, 0)$ for AA-stacking with $\varphi = 2\pi/3$. 
When $\bm{\tau} = (0, a/\sqrt{3})$ for AB-stacking the $T^0$ matrix remains the same but 
$T^{\pm}$ acquires a phase factor as follows,
\begin{equation}
T^{\pm} = \left(\begin{array}{cc}
\omega' e^{\mp i\varphi} & \omega e^{\pm i\varphi}\\
\omega  & \omega' e^{\mp i\varphi}\\
\end{array}\right),
\end{equation}
and when $\bm{\tau} = (0, 2a/\sqrt{3})$ or, equivalently, $\bm{\tau} = (0, -a/\sqrt{3})$
for BA stacking the matrix is conjugate transposed. 
The interlayer tunneling elements use the polynomial parametrization of Ref.~\cite{Chebrolu2019}
relating inter- and intra-sublattice hopping terms $\omega' = A\omega^2 + B\omega + C$ where $A = -0.5506$, $B = 1.036$, and $C = -0.02245$ that is fitted to the exact exchange and random phase approximation (EXX+RPA) interlayer energy minima and local density approximation (LDA) interlayer tunneling, 
and leads to different $\omega' = 0.0939$~eV and $\omega  = 0.12$~eV when effective out of plane relaxations are considered. Equal tunneling parameters $\omega' = \omega = 0.12$~eV correspond to the rigid model in the absence of relaxations, which are consistent with the LDA values for the $t_{1} \simeq 3 \omega = 0.36$~eV perpendicular interlayer tunneling term in an AB stacked bilayer~\cite{accuratebilayer}.

In tTG with aligned top and bottom layers we have two moire interfaces with the same moire length 
$L_M = a/ (2\sin{(\theta/2)})$. 
The magic angle given as $\theta \simeq 1.5^\circ \simeq \sqrt{2} \times 1.06^{\circ}$ is 
enlarged with respect to the tBG value by a factor of $\sqrt{2}$ following the renomalization of the interlayer tunneling strength when we decompose the interaction of the outer layers Dirac Hamiltonian with the middle 
layer~\cite{Khalaf2019, Carr2019a, Li2019}. 
The ${\bm \tau}$ top layer sliding vector with respect to bottom layer is a control knob that alters the 
electronic structure of our system. 
For most cases we choose ${\bm \tau}_{\rm AA} = (0,0)$ where top and bottom layers are exactly on top of
each other, ${\bm \tau}_{\rm AB} = (0,a/\sqrt{3})$ where the top layer has a Bernal stacking-like displacement, 
and the intermediate saddle point stacking ${\bm \tau}_{\rm SP} = (0, a \sqrt{3}/2)$ is chosen as the representative broken 
rotational symmetry system leading to clearest strip patterns. 
We interchangeably refer to the AA and AB stacking of the top-bottom layers in tTG
with the sliding vectors ${\bm \tau}_{\rm AA}$ and ${\bm \tau}_{\rm AB}$, see Fig.~\ref{fig1}.
The local AAB-stacking is generated by sliding the top layer by $\bm{\tau}_{\rm AB}$ from  
the AAA-stacking where all three layers are exactly aligned on top of each other.
Because the stacking sliding geometry of the middle layer does not alter the resulting band structure after it is twisted
we use the bottom and top layer stacking labels to classify the different systems.
When we twist the middle layer by the magic angle $\theta \simeq 1.5^\circ$, see Fig.~\ref{fig1},   
we can identify two overlaid patterns with equal period 
$\sim 9.59$~nm where a finite $\bm{\tau}$ introduces changes in the local stacking maps. 
The black letters represent on top of the moire patterns the local stacking geometries 
as shown schematically in the second row.
The AA-tTG has mirror symmetry with respect to the middle layer as illustrated from the 
AAA, ABA, and BAB local stacking
configurations, while mirror symmetry is broken in AB-tTG but an inversion center is present
for all twist angles, preserving in both cases the triangular rotational symmetry of the moire patterns.
All these symmetries are broken for intermediate $\bm{\tau}$ vectors away from the symmetric stacking configurations,
and this is illustrated for a large commensurate twist angle $\theta = 21.8^\circ$ and three 
different top layer sliding $\bm{\tau}$ vectors in the third row of Fig.~\ref{fig1}.

\section{ENERGY BANDS}\label{R1}
The electronic structure of tTG strongly depends on system parameters such as twist angle $\theta$, 
the interlayer potential difference $\Delta$, and top layer sliding vector $\bm{\tau}$. 
Here, our bandwidth phase diagram analysis for tTG shows that in addition 
to the magic angle $\theta \sim 1.5^{\circ}$
the narrowest $W$ are found for zero or moderate values of $\Delta$ 
at a smaller twist angle near $\theta \sim 1.2^{\circ}$ for $\bm{\tau}_{AA}$, 
and near $\theta \sim 1.4^{\circ}$ for $\bm{\tau}_{AB}$, 
and for all considered $\Delta$ and $\bm{\tau}$ when $\theta \lesssim 0.6^{\circ}$.
%
%
%
%
Sample electronic structure surface plots and contours are shown in Fig.~\ref{fig2} 
for $\theta = 1.5^{\circ}$ near the magic angle and in Figs.~\ref{fig3} and \ref{fig3b}
we present continuous sweep phase diagrams of electronic structure features
in the parameter space of $\theta$, $\Delta$ and $\bm{\tau}$. 
In this work we focus our attention on systems with $\omega' \neq \omega$ with 
$\omega' = 0.0939$~eV and $\omega = 0.12$~eV in Eq.~(\ref{tunneling}) that accounts for out of plane 
relaxations that gaps the Dirac cones at $\tilde{\Gamma}$ for AA and at $\tilde{K}$ for AB.
%

%
We begin by illustrating in Fig.~\ref{fig2} the 
impact of the stacking type $\bm{\tau}$ in the bandwidth $W$ corresponding to the
valence and conduction low energy bands that give rise to the progressively increasing
sequence $W(\bm{\tau}_{\rm AA}) \lesssim 2$~meV,
$W(\bm{\tau}_{\rm AB}) \lesssim 15$~meV and $W(\bm{\tau}_{\rm SP}) \lesssim 40$~meV 
as we depart from the initial $\bm{\tau}_{\rm AA} = (0,0)$ stacking geometry for 
$\theta=1.5^{\circ}$ and show in the figures for $\theta = 1.2^{\circ}, \, 1.4^{\circ}$.
We will show that a finite interlayer potential difference $\Delta$ alters the bandwidths giving rise to 
a roughly linear increase $W \propto \Delta $ near the magic angle for AA and a 
non-monotonic behavior for AB and SP stackings.
For the AA case a finite $\Delta$ shifts the band touching point at $\tilde{K}$ to proportionally 
higher positive and negative energy values without opening a primary band gap $\delta_{p}$ 
nor secondary gaps $\delta_s$ in both the valence and conduction bands\cite{Carr2019a, tb_tTG}, 
while for AB we have positive $\delta_{p}$ and $\delta_s$ gaps~\cite{park2020tunable, hao2020electric}.
The opening of the band gaps and subsequent isolations possible for AB systems leads 
to low energy bands with well defined valley Chern numbers depending on $\theta$ and $\Delta$ values
in contrast to AA bands that remain metallic.

The bandwidth $W$ for conduction and valence bands and the associated primary $\delta_p$ gap 
and secondary gap $\delta_s$ for different system parameters are illustrated in Fig.~\ref{fig3} for continuous 
variations of $\theta$ and $\Delta$ for select $\bm{\tau}$ values, and for continuous $\bm{\tau}$ for 
a fixed $\theta = 1.5^{\circ}$ and select $\Delta$ values in Fig.~\ref{fig3b}.
The bandwidth phase diagrams and gaps are strongly affected by $\bm{\tau}$
and the results in Fig.~\ref{fig3} shows that the electron hole asymmetry is generally 
weaker in our tTG models where we do not incorporate interactions with the substrate~\cite{shi2021,jiseon2021}
nor the remote hopping terms included in a Bernal stacked bilayer graphene~\cite{mccann2006,accuratebilayer}.
This is manifested in the closely resembling behavior of the different 
$W$, $\delta_p$, $\delta_s$ phase diagrams for the conduction and valence bands.
As we just noted, for $\bm{\tau}_{\rm AA}$ we generally find metallic bands that have 
narrowest bandwidths near the magic angle $\theta \simeq 1.5^{\circ}$, 
a slightly lower $\theta \simeq 1.2^{\circ}$, and $\theta \lesssim 0.5^{\circ}$.
The presence of interlayer potential differences $\Delta$ introduces a mild almost linear increase in the 
bandwidths near $\theta$ that follows approximately the relation $W \simeq 0.4 \, \Delta$, 
indicating that narrowest bands are expected when there are no displacement electric fields. 
The bandwidths remain consistently narrow $W \lesssim 10$~meV
for all considered values of $\Delta$ in the small twist angle regime when $\theta \lesssim 0.6^{\circ}$.
The situation is different for $\bm{\tau}_{\rm AB}$ where isolated bands
can be found in the presence of a finite $\Delta$ between a wider $1^{\circ} \sim 1.7^{\circ}$ twist angles 
range and at islands near $\sim 0.6^{\circ}$ for sufficiently large $\Delta$, 
and near $\sim0.4^{\circ}$ for all values of $\Delta$.
The narrowest bandwidth regions are found in the vicinity of $\theta \simeq 1.4^{\circ}$ slightly below
the magic angle and $\theta \lesssim 0.6^{\circ}$. 
A finite $\Delta$ at $\theta$ shows a non-monotonic behavior in $W$ 
reducing the bandwidth with $\Delta$ before it eventually 
recovers the almost linear relationship $W \simeq 0.25 \, \Delta$ beyond $\Delta \simeq 0.04$~eV where
the secondary band gaps start to decrease. 
Similar to $\bm{\tau}_{\rm AA}$, the bandwidths remain consistently narrow $W \sim 10$~meV
for all considered values of $\Delta$ when $\theta \lesssim 0.6^{\circ}$.
Finally, for a third sliding vector $\bm{\tau}_{\rm SP}$ corresponding to a saddle point stacking
the bandwidths remain practically constant with a value on the order 
of $\sim$40~meV for twist angles between  $1.2^{\circ} \sim 1.7^{\circ}$,
while narrowest bandwidths are expected for small twist angles
$\theta \lesssim 0.5^{\circ}$ in the range of explored $\Delta$ 
values to up to 0.1~eV like in the other $\bm{\tau}$ configurations.

\begin{figure*}
\begin{center}
\includegraphics[width=18cm]{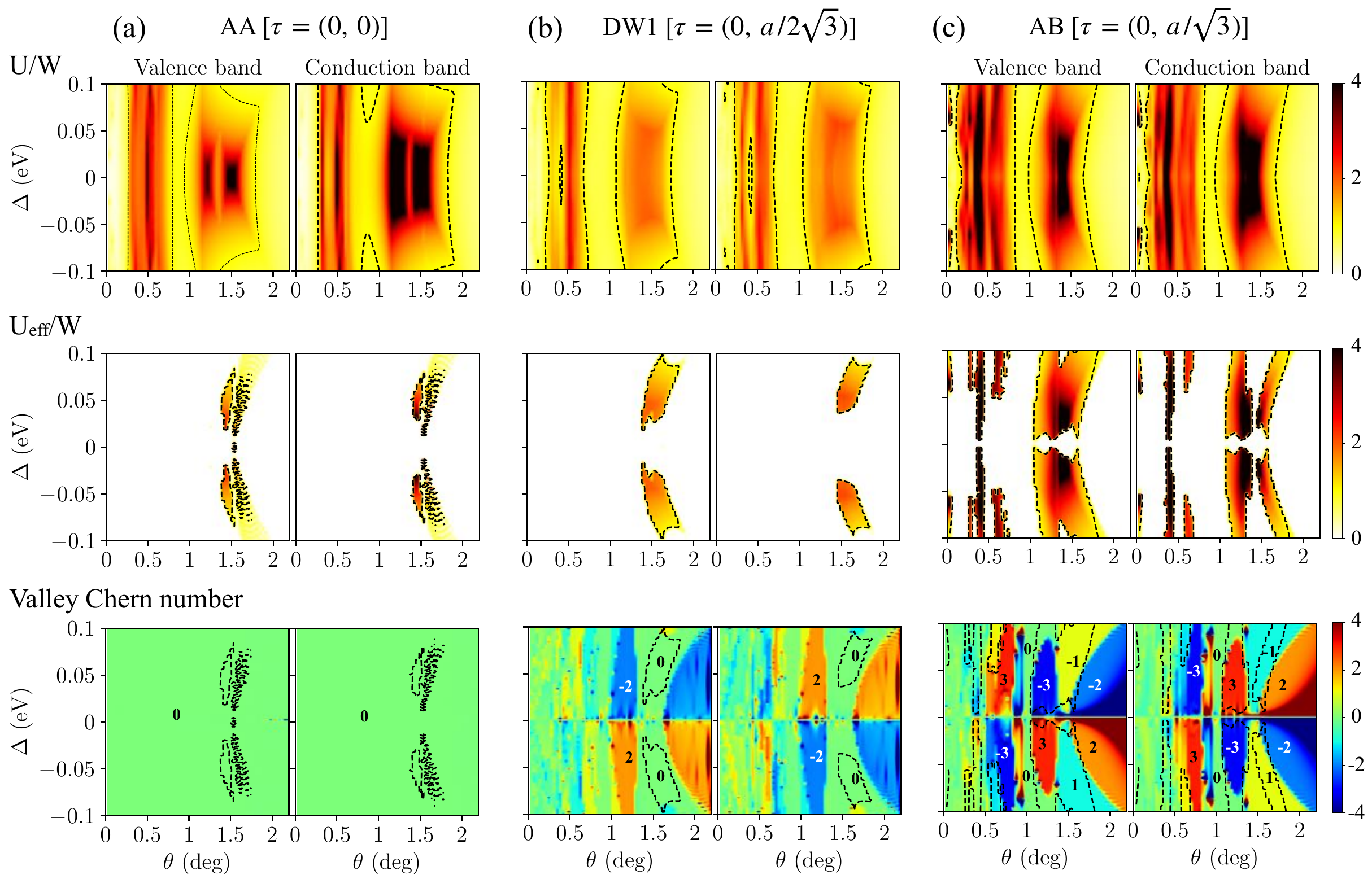}
\end{center}
\caption{ (Color online) The ratio of $\bm{\tau}$ the bare Coulomb interaction U to the bandwidth W (upper),  the ratio of the screened Coulomb interaction U$_\mathrm{eff}$ to the bandwidth W (middle), and the valley Chern numbers (lower) for the valence and conduction bands for (a) AA [$\bm{\tau} = (0, 0)$] and AB-starting stacking [$\bm{\tau} = (0, a/\sqrt{3}$)] in tTG for the $\omega' \neq \omega$ model 
as a function of gate voltage $\Delta$ and twist angle $\theta$ at the magic angle $\theta = 1.5 ^\circ$. 
	}\label{fig4}
\end{figure*}

\begin{figure*}
\begin{center}
\includegraphics[width=18cm]{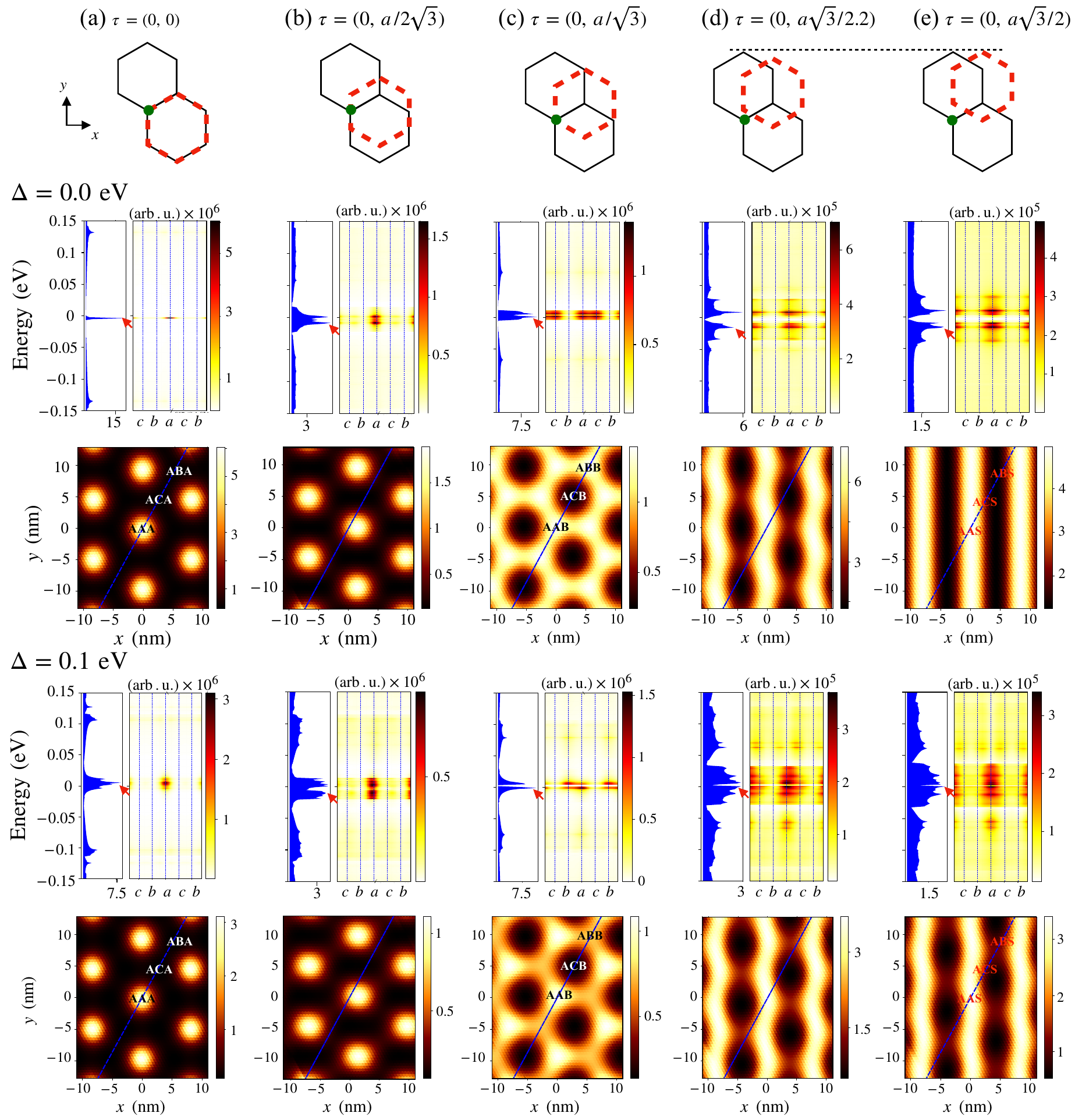}
\end{center}
\caption{ (Color online) 
Electronic density of states (DOS) and the corresponding local density of states (LDOS) 
of tTG for the $\omega' \neq \omega$ model with gate voltage $\Delta = 0.0$ eV and $0.1$ eV for five different starting stackings, 
(a) $\bm{\tau} = (0, ~0)$ [AA], (b) $(0, a/2\sqrt{3})$, (c) $(0,  a/\sqrt{3})$ [AB],  
(d) $(0, a \sqrt{3}/2.2)$, and (e) $(0, a\sqrt{3}/2)$ [AS].  
	The first row: Schematic diagrams for each starting stacking with the slid top graphene layer denoted by a red dashed hexagon and the rotation axis indicated by a green filled circle. 
	Second \& fourth rows: DOS (left) and LDOS (right) along a blue dashed straight line
	($c-b-a-c-b$) indicated on the third and fifth rows. 
	($a$, $b$, $c$) letters label respectively the (AA, AB, AC) local stacking between the bottom 
	and middle layer atoms and they are 
	equal to (AAA, ABA, ACA) for $\bm{\tau}_{\rm AA}$ top-bottom layers stacking in (a), 
	(AAB, ABB, ACB) for $\bm{\tau}_{\rm AB}$ stacking in (c), (AAS, ABS, ACS) for 
	$\bm{\tau}_{\rm AS}$ stacking in (e).
	Third \& fifth rows: LDOS in a two-dimensional real space at the van Hove singularity (vHs) (lower row).
	}\label{fig5}
\end{figure*}

\section{EFFECTIVE COULOMB INTERACTION and VALLEY CHERN NUMBERS}\label{R2}
In this section we provide measures on the relative dominance of the Coulomb interaction energies
versus the bandwidth $W$ by calculating the ratios $U/W$ of the bare Coulomb energy versus bandwidth, 
and the screened effective Coulomb energy versus bandwidth $U_\textrm{eff}/W$ that provides 
a more reliable measure for the onset of gaps and insulating phases when the bands are not overlapping. 
Typically we consider to be in the strong correlation regime when these ratios are larger than 1. 
The effective screened Coulomb energy is given by~\cite{Chebrolu2019} 
\begin{equation}
U_\textrm{eff} = \frac{e^2}{4 \pi \epsilon_r \epsilon_0 l_M} \exp{(-l_M/ \lambda_D)},
\end{equation}
where the moire length is given by $l_M = a/(2 \sin(\theta/2)) \simeq a/\theta$.
The effective screening Debye length $\lambda_D$ is expressed as 
$\lambda_D = 2 \epsilon_0 / e^2 D(\delta_p, \delta_s)$ where $D(\delta_p, \delta_s)$ is the 
two-dimensional DOS defined as
\begin{equation}
D(\delta_p, \delta_s) = 4\frac{\vert \delta_p \vert u(-\delta_p) + \vert \delta_s \vert u(-\delta_s)}{W^2 A_M},
\end{equation}
where $u(x)$ is the Heaviside step function. 
Thus, $D(\delta_p, \delta_s)$ is proportional to the bands overlap $|\delta_{p(s)}|$ 
represented by negative gap values $\delta_p < 0$ ($\delta_s < 0$). 
Here, $A_M = \sqrt{3} \, l^2_M/2$ is the area of a moire unit cell in real space, 
and we use $\epsilon_r = 4$ for the dielectric constant of graphene.

The first two rows in Fig.~\ref{fig4} show $U/W$ and $U_{\rm eff}/W$ as a function of twist angle and interlayer
potential difference $\Delta$ for the valence and conduction flat bands, 
which generally show a weak electron-hole asymmetry regardless of the 
different top-botoom layer sliding geometries considered, namely 
$\bm{\tau}_{\rm AA} = (0,0)$, the intermediate $\bm{\tau}_{\rm DW1} = (0, a/ (2\sqrt{3}))$, 
and $\bm{\tau}_{\rm AB} = (0, a/\sqrt{3})$.

The first row showing $U/W$ plots resembles the bandwidth $W$ phase diagram in Fig.~\ref{fig3}
manifesting a strong dependence with respect to $\bm{\tau}$.
We indicate the contours of $U/W = 1$ with black dotted lines to help distinguish 
the regions where we expect strong correlations. 
For $\bm{\tau}_{\rm AA}$ the large Coulomb energy regions are found at the aforementioned 
bandwidth minima angles of $\theta \simeq 1.5^{\circ}, 1.2^{\circ}$ and for angles below $\sim 0.6^{\circ}$.
Sliding to an intermediate stacking $\bm{\tau}_{\rm DW1}$ has the effect of reducing the overall 
strength of the $U/W$ ratio seen in $\bm{\tau}_{\rm AA}$, 
and further sliding until $\bm{\tau}_{\rm AB}$ achieves a wider region of large $U/W$ ratios 
in the parameter space of $\theta$ and $\Delta$
where peak maxima are shifted to a lower $\theta \simeq 1.4^{\circ}$ and for angles below $\sim 0.8^{\circ}$.
%
%
The second row showing $U_{\rm eff}/W$ includes suppression of the Coulomb energy due to screening effects
proportional to the overlap of the flat bands with the neighboring bands. 
Similar to the first row, we indicate the contours of $U_\textrm{eff}/W = 1$ with black dotted lines.
This quantity allows to define the regions where the bands are isolated and 
we have a higher likelihood of developing insulating gapped phases. 
For all stacking geometries considered we observe that twist angles around $\theta \sim 1.5^{\circ}$
within $\pm 0.2^{\circ}$ can develop $U_{\rm eff}/W \gtrsim 1$ regions when we add a sufficiently large $\Delta$.
%

The valley Chern numbers corresponding to the flat bands are represented in the third row of Fig.~\ref{fig4}
and they will be well defined when the band are not crossing each other. 
%
The valley Chern number of the $n^{\rm th}$ energy band $C_n$ is defined as
\begin{equation}
C_n = \frac{1}{2 \pi} \int_\textrm{mBZ} d^2 \bm{k} ~ \Omega_n (\bm{k}),
\end{equation}
where $\Omega_n (\bm{k})$ is the Berry curvature given by Ref.~\cite{Xiao2010b} as follows,
\begin{equation}
\Omega_n (\bm{k}) = -2 \sum_{n' \neq n} \textrm{Im} \Bigg[ \frac{\langle n \vert \frac{\partial H}{\partial k_x} \vert n' \rangle \langle n' \vert \frac{\partial H}{\partial k_y} \vert n \rangle}{(E_{n'} - E_n)^2} \Bigg].
\label{Bcurv}
\end{equation}
The $K$ valley Chern numbers of both the valence and the conduction low energy bands are shown 
in the lower row in Fig.~\ref{fig4}. 
For $\bm{\tau}_{\rm AA}$ they are found to be topologically trivial for finite electric fields
and twist angles less than 2.5$^\circ$, while for twist angles larger than 3$^\circ$ 
the valley Chern numbers are not well-defined due to the strongly metallic character of the system (not shown). 
On the other hand, the valley Chern numbers of $\bm{\tau}_{\rm DW1}$ and $\bm{\tau}_{\rm AB}$ cases 
show diverse topologically nontrivial phases. 
One general observation is that the valley Chern number signs can be reversed with the perpendicular
electric field direction, and the valley Chern numbers of the valence and conduction bands are 
opposite to each other adding up to a zero sum.
For $\bm{\tau}_{\rm DW1}$ we expect trivial Chern numbers for 
strong correlation regions for $\theta \sim 1.5^{\circ}$ and finite $\Delta$, and $\pm 2$ valley Chern numbers
for twist angles that are below and above in twist angle. 
For $\bm{\tau}_{\rm AB}$ the parameter space of isolated bands and finite valley Chern numbers are expanded 
thanks to the easier opening of band gaps $\delta_{p}$ and $\delta_{s}$.
Near the magic angle $\theta \sim 1.5^{\circ}$ the valley Chern numbers are finite $C = \pm 1$,
becoming $C = \pm 3$ for smaller angles, and becoming $C = \pm 2$ larger angles.
The valley Chern numbers become again mostly trivial 
for the saddle point AS stacking with $\bm{\tau}_{\rm SP} = (0, a_{\rm G} \sqrt{3}/2)$ 
when the system behaves like a metal.
Our calculations show that a variety of finite valley Chern numbers can be tailored 
depending on the specific top-bottom layer sliding configuration $\bm{\tau}$.

\begin{figure*}
\begin{center}
\includegraphics[width=0.98\textwidth]{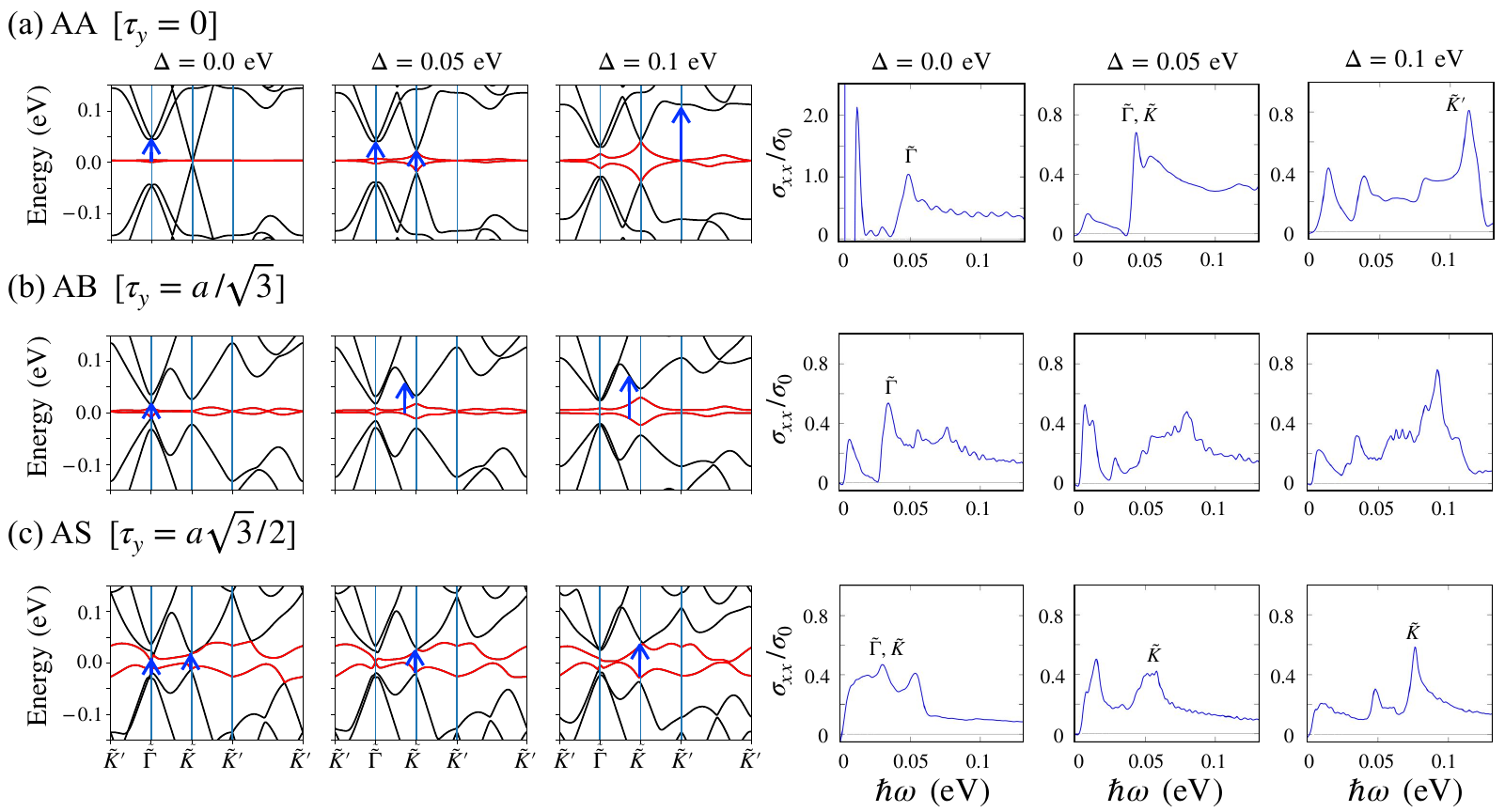}
\end{center}
\caption{ 
(Color online) Electronic band structures of tTG at its magic angle $\theta = 1.5^{\circ}$ 
and the corresponding longitudinal linear optical conductivity $\sigma_{xx}$ 
at zero chemical potential for (a) AA, (b) AB, (c) AS-starting stackings for different displacement fields $\Delta = 0.0$, $0.05$, $0.1$~eV.  
Each prominent peak in the optical conductivity is indicated by a blue diamond and the contributions 
for the each transition are denoted by the blue arrows in the band structures and highlighted in its mBZ.  
}\label{fig6}
\end{figure*}

\section{LOCAL DENSITY OF STATES}\label{R3} 
We have just noted that the electronic structure undergo important changes depending on the 
sliding vector $\bm{\tau}$ following variations of the real space moire patterns,
which in turn impacts the local density of states (LDOS) associated to the nearly flat bands.
Here we show that LDOS maxima locations in tTG follow closely the same
rule of thumb applicable in tBG that concentrates the charge at the AA local stacking regions of a tBG interface. 
Because in tTG we have two tBG interfaces the AA local stacking centers at each interface
will distribute in different manners depending on the $\bm{\tau}$ vector, 
as we illustrate in Fig.~\ref{fig5} for five selected cases of top layer sliding in the $y$-direction. 

%
For the $\bm{\tau}_{\rm AA}$ case the reinforced LDOS profiles 
at the AAA local stacking regions give rise to a triangular lattice much like what we find in tBG.
The LDOS patterns progressively split into two displaced triangular lattices 
and thus breaks the triangular rotational symmetry 
as we introduce a small sliding in the top graphene layer by a $\bm{\tau} = (0, a/2\sqrt{3})$ 
vector along the $y$-direction in Fig.~\ref{fig5}(b).
The triangular rotational symmetry is recovered for $\bm{\tau}_{\rm AB} = (0, a/\sqrt{3})$, see Fig.~\ref{fig5}(c),
and the LDOS maxima forming a honeycomb lattice
consisting of AA local interfaces between bottom-middle and middle-top layers.
When we continue sliding the top layer further the double moire pattern 
start forming stripe shapes for the LDOS in Fig.~\ref{fig5}(d)(e) which break maximally the rotational symmetry 
at the so called saddle point $\bm{\tau}_{\rm SP} = (0, a \sqrt{3}/2)$.
These LDOS charge anisotropy patterns in Fig.~\ref{fig5}(d) and (e) resemble the scanning 
tunneling spectroscopy (STS) results in Ref.~\cite{Zuo2018} of twisted trilayer graphene.

Application of a positive interlayer potential difference $\Delta$ has the effect of redistributing the carrier densities
of the valence bands towards the bottom-middle interface as it tends to lower the bottom layer energy
following the definition in Eq.~(\ref{eq1}). 
Thus for the honeycomb pattern resulting from the $\bm{\tau}_{\rm AB} $ stacking we see a 
brightened bottom-middle interface and dimmed middle-top interface upon application of $\Delta$.
For similar reasons, a finite $\Delta$ distorts the straight stripe patterns seen in Fig.~\ref{fig5}(e) 
turning them into snake like shapes by populating the bottom interface and depleting the top interface 
charge densities near the respective AA stacking regions. 
In general an interlayer potential difference has the effect of broadening  
the DOS in energy pushing the states to higher energy values away from neutrality,
and for the $\bm{\tau}_{\rm AB}$ stacking case we see the opening of  
a band gap at charge neutrality from the DOS profile.


\section{LONGITUDINAL OPTICAL CONDUCTIVITY}\label{R4}


In this section, we present the numerical analysis of the longitudinal linear optical conductivity 
of tTG at the magic twist angle $\theta =1.5^\circ$ for select values of the displacement 
field for the three 
stacking arrangements $\bm{\tau}_{\rm AA}$, $\bm{\tau}_{\rm AB}$ and $\bm{\tau}_{\rm AS}$
in Figs.~\ref{fig6}(a), (b), and (c), respectively. 
The real part of the longitudinal linear optical conductivity is given by~\cite{ando2002dynamical, gusynin2006unusual, gusynin2007anomalous, falkovsky2007space, min2009}

\begin{widetext}
\begin{equation}
Re[\sigma_{xx}(\omega)]/\sigma_0 = \frac{16}{\omega} \int \frac{d^2 \bm{k}}{(2\pi)^2}\sum_{i,j} [f(\epsilon_{\bm{k},i})-f(\epsilon_{\bm{k}, j})] \vert \langle \bm{k}, i \vert  J_x \vert \bm{k}, j\rangle \vert^2 \delta[\omega + (\epsilon_{\bm{k}, j} - \epsilon_{\bm{k},i})/\hbar],
\end{equation}
\end{widetext}
where $J_\alpha = - \partial H / \partial k_\alpha$ is the general current operator, $f(\epsilon)$ is the Fermi-Dirac distribution function,
$\epsilon_{\bm{k}, i}$ is the $i$th eigen-energy at $\bm{k} = (k_x,~ k_y)$, and $\sigma_0 = \pi e^2/2h$ is the universal optical conductivity of the single layer of graphene.

In Fig.~\ref{fig6}, we illustrate the energy bands for select band structures together with the 
real part of the linear optical conductivity at zero chemical potential. 
We have not considered the Drude term in order to present more clearly the 
contributions of interband optical transitions. 
In the band structure figures, the lowest electron- and hole-bands are highlighted by the red lines. 
The corresponding real part of the normalized linear optical conductivity in the longitudinal 
direction $\sigma_{xx}/\sigma_0$ are juxtaposed together with the mBZ maps on the right panel. 
We stressed the prominent contributions in the conductivity denoted by the blue diamonds 
and investigated the locations of the optical transitions in the mBZ maps. 

The locations of each transition peak in momentum space are also illustrated in the band structure figures by the blue arrows. 
For $\bm{\tau}_{\rm AA}$ stacking cases, the largest contributions of the optical transitions occur at $\tilde{\Gamma}$ 
or $\tilde{K}$ points when the displacement fields are $\Delta$ =  $0.0$, $0.05$~eV. 
On the other hand, the biggest portion of the transition takes place at $\tilde{K}'$ when $\Delta$ =  $0.10$~eV.  
In $\bm{\tau}_{\rm AB}$ stacking, the biggest contribution of the transitions happens at $\tilde{\Gamma}$ when 
$\Delta$ = $0$~eV. On the other hand, the transitions mostly occur at an intermediate point away from 
the high-symmetry point for $\Delta$ = $0.05$, $0.10$~eV. 
For the $\bm{\tau}_{\rm AS}$-stacking case, the largest optical contributions are mostly coming from $\tilde{K}$ point and it is 
noteworthy that the contribution in the mBZ map is anisotropic, which reflects the triangular rotational symmetry
in keeping with the anisotropic Fermi surface as well as the real-space stripe patterns.

\section{SUMMARY}\label{Summary}
Trilayer graphene with middle layer twist (tTG) gives rise to the simplest form of commensurate 
double moire pattern formed by two twisted graphene interfaces and 
has become a new system of interest following recent observations of superconductivity with higher
critical temperatures than in twisted bilayer graphene (tBG).
We have presented a detailed electronic structure calculations and associated 
phase diagrams for the bandwidth, gaps and valley Chern numbers
for continuous variations of the twist angle $\theta$ and interlayer potential difference $\Delta$
for selected $\bm{\tau}$ top-bottom layer sliding vectors.
We have aimed at providing a more comprehensive 
description of the system behavior in a wider range of system parameters than in earlier work
to predict new system parameters where strong correlations and finite valley Chern numbers are expected,
and paid particular attention to the role of the $\bm{\tau}$ interlayer sliding 
that can either preserve or break the triangular rotational symmetry to create anisotropic strip patterns.
While the bandwidths of the low energy states generally follow the
$W(\bm{\tau}_{\rm AA}) < W(\bm{\tau}_{\rm AB}) < W(\bm{\tau}_{\rm SP})$ sequence
they are modified by $\Delta$ which alters the twist angle dependent bandwidth phase diagram.
Our calculations predict narrowest bandwidths on the order of $\sim$10~meV 
around $\theta \simeq 1.5^{\circ}$ and $1.2^{\circ}$ in the limit of small $\Delta$ for $\bm{\tau}_{\rm AA}$ stacking, 
and around $\theta \simeq 1.4^{\circ}$ for $\bm{\tau}_{\rm AB}$ stacking.
Application of a finite $\Delta$ generally widens the bandwidth of the low energy flat bands
and in the case of $\bm{\tau}_{\rm AB}$ the low energy bands can be isolated 
to generate finite valley Chern numbers in a wide range of twist angles $\theta$ 
and interlayer potential difference $\Delta$.
We have also analyzed the impact of stacking and electric fields in the local density of states (LDOS) 
maps that can be measured through scanning tunneling probes, 
and showed that the anisotropic stripe patterns can be maximized when the top-bottom layers 
have a saddle point stacking geometry. 
The specific stacking vector $\bm{\tau}$ favored in the system might be modifiable 
through different device preparation conditions, for example in the presence of strains
introduced by boundary condition stresses, that would in turn lead to observable changes 
in charge transport or through optical experiments. 
The linear optical conductivity calculations we have carried out provide information about 
the changes expected in the interband
transition peaks that can be introduced by varying the system parameters 
and suggests its usefulness as a system characterization tool.

\begin{acknowledgments}
We gratefully acknowledge Y. J. Park's help in the preparation of some figures. 
This work was supported by Samsung Science and Technology Foundation under 
project no.~SSTF-BA1802-06 for J. S., 
the Korean National Research Foundation grants NRF-2020R1A2C3009142 for B. L. C., 
and NRF-2020R1A5A1016518 for J. J. 
We acknowledge computational support from KISTI through Grant KSC-2020-CRE-0072.
\end{acknowledgments}

\begin{appendix}

\section*{Appendix. Band structures for select cases}\label{AppendixA}
\begin{figure*}
\begin{center}
\includegraphics[width=0.8\textwidth]{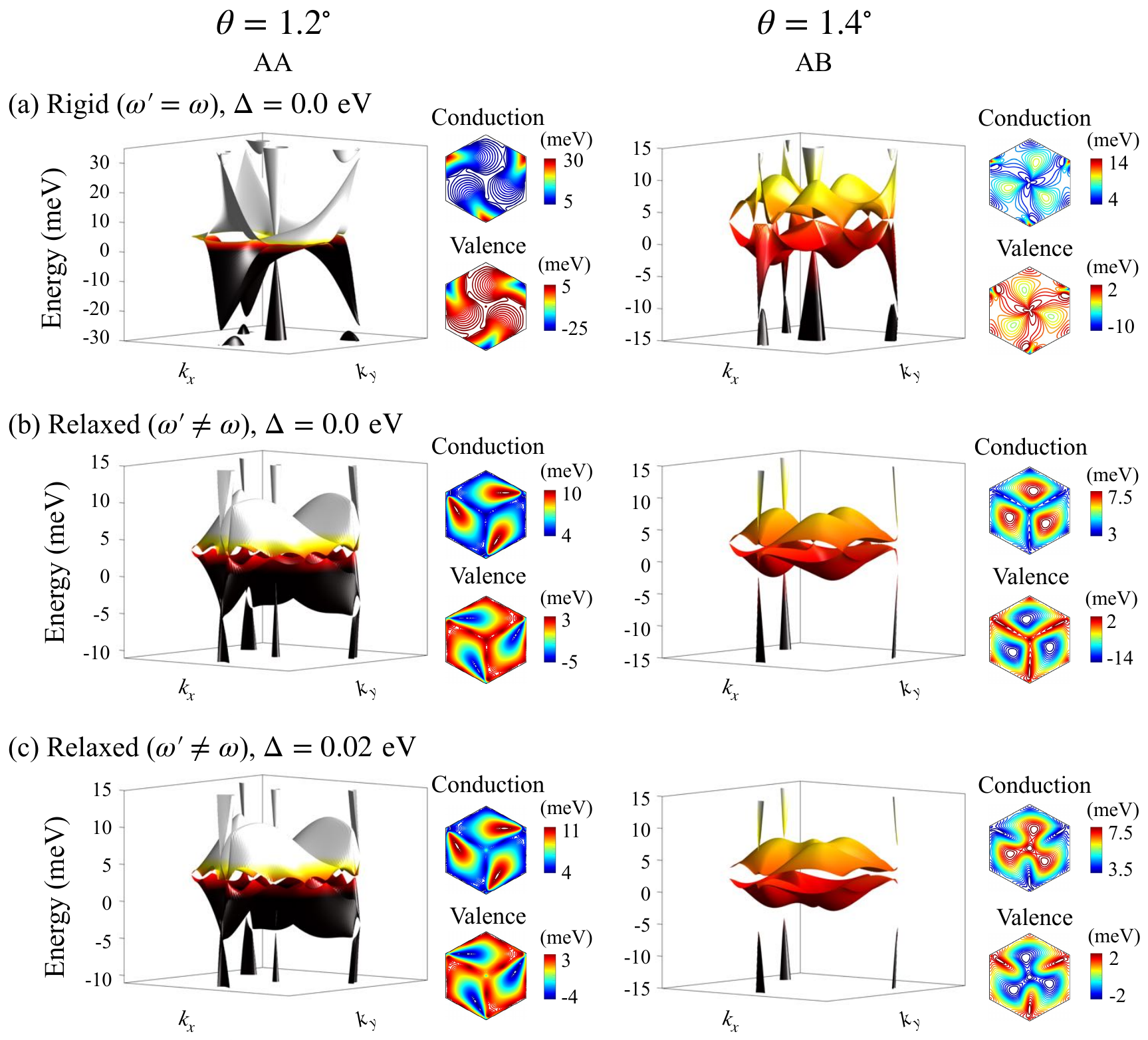}
\end{center}
\caption{ 
(Color online) Electronic band structures of tTG with $\tau_{AA}$ at $\theta = 1.2^\circ$ and with $\tau_{AB}$ at $\theta = 1.4^\circ$ for (a) rigid lattice ($\omega' = \omega = 0.12$~eV), 
(b) unequal tunneling $\omega' = 0.0939$~eV and $\omega = 0.12$~eV that flattens the bands by suppressing the bandwidth at the moire Brillouin zone corners,
and (c) with the finite interlayer potential difference $\Delta = 0.02$ eV. 
The Fermi surface contours of the conduction and valence bands are plotted together with the band structures. 
}\label{figA1}
\end{figure*}

In this appendix we additionally present in Fig.~\ref{figA1} the band structures for the cases of having the local minima in the conduction and valence bandwidths for $1.2^{\circ}$ and $1.4^{\circ}$ in addition to the $1.5^{\circ}$ case that we showed in Fig.~\ref{fig3}(a) and (b). 
When the top and the bottom layers have a relative displacement by $\tau_{AA}$,  the valence bandwidth at $\theta \sim 1.2^\circ$ has the local minimum $W \sim 8$~meV, and the conduction bandwidth at $\theta \sim 1.2^\circ$ has the local minimum $W \sim 6$~meV. For the top-bottom layer displacement $\tau_{AB}$, both valence and conduction bandwidths have the local minima $W \sim 5$~meV at $\theta \sim 1.4^\circ$. The corresponding band structures for the rigid ($\omega' = \omega = 0.12$~eV), 
the out-of-plane relaxed lattice ($\omega' = 0.0939$~eV and $\omega = 0.12$~eV), 
and with finite displacement field $\Delta = 0.025$ eV are shown on the left (right) column for $\tau_{AA}$ ($\tau_{AB}$) 
in Fig.\ref{figA1}(a), (b), and (c), respectively. We note that unequal interlayer tunneling $\omega \neq \omega'$ helps to flatten the
low energy bands by reducing the band dispersion at the moire Brillouin zone corners.

\end{appendix}

\bibliography{tTG_sliding}

\end{document}